# THREE DIMENSIONAL BRIGHT VORTEX SOLITON

Lubomir M. Kovachev

Institute of Electronics, Bulgarian Academy of Sciences

Tzarigradsko chaussee 72,1784 Sofia, Bulgaria

Abstract. The scalar theory of the self-focusing of an optical beam is not valid for very narrow beam. Using the method of separation of variables, we developed a vector nonparaxial theory from the nonlinear wave equations in strong field approximation. We found that exact localized vortex soliton solution exist for the nonlinear 3D+1vector wave equation and for the 3D+1 vector nonlinear Schrodinger equation. This method is applicable for angular functions, which satisfied additional conditions. It is shown that exact vortex solitary wave exist only for solution with eigenrotation momentum L=1.

PACS number 42.81.Dp

1. Introduction.

The early theories [1-4] of propagation of an optical beam in nonlinear cubic media was based on investigation of scalar paraxial wave equations and predicted a self -focusing effect. In the last thirty ears, significant results have been obtained in the investigation of localized optical pulses in 1D+1 dimension approximation. The existence was demonstrated of stable localized wave packets, solitons, in optical fibers and slab wave-guides. Recently, there has been much interest [5-8] in the investigation of vortex solitary waves in nonlinear media. The study of the optical vortices was made numerically by using scalar paraxial wave equations. As was pointed out in [9,10], the scalar theory is not valid for a very narrow beam, and a correct description of the beam behavior requires a vector analysis. Using an order of magnitude analysis method, it is no hard to see that all these results were obtained for wave packets in a weak nonlinear media. One quantitative criterion for weak nonlinearly, is the dimensionless magnitude to be in order of:

$$\frac{\tilde{n}_2}{n_0^2}|E_0|^2 \approx \mu^2, \text{where } \mu \ll 1.$$

Presently, there are lasers system producing localized optical wave packets with magnitude in the order of:

$$\frac{\tilde{n}_2}{n_0^2}|E_0|^2 \approx 1 \qquad (1)$$

in suitable nonlinear media. For this media condition (1) is the strong field approximation. In this paper we investigate a vector wave equation, derived from the Maxwell equations when the condition (1) is satisfied.

2. The equation and order of magnitude analysis.

We started from the nonlinear wave equations derived from the Maxwell's equations of nonlinear cubic media:

$$-\Delta\vec{E} + \nabla(\nabla\cdot\vec{E}) + \frac{n_0^2}{c^2}\frac{\partial^2\vec{E}}{\partial t^2} = -\frac{n_0^2}{c^2}\frac{\partial^2\vec{P}_{nl}}{\partial t^2}. \qquad (2)$$

$$\vec{P}_{nl} = \frac{\tilde{n}_2}{n_0^2}|\vec{E}|^2\vec{E}, \qquad (3)$$

where $n_0$ is the linear refractive index and $\tilde{n}_2$ is the Kerr coefficient.

As was pointed out in [10], for an isotropic media, the frequency-degenerate nonlinear polarization $\vec{P}_{nl}$ of an arbitrary electric field $\vec{E}$ is always parallel to $\vec{E}$ and led to the fact that $\chi^{(3)}$ is a scalar. If we neglect the term $\nabla(\nabla \cdot \vec{E})$, we obtain the next vector wave equation from (2) and (3):

$$-\Delta\vec{E} + \frac{n_0^2}{c^2}\frac{\partial^2 \vec{E}}{\partial t^2} = -\frac{n_0^2}{c^2}\frac{\partial^2 \vec{P}_{nl}}{\partial t^2} \qquad (4)$$

It is well known that a spectral limited localized wave packet satisfies the conditions:

$$\Delta\omega \approx \mu.\omega_0; \Delta k \approx \mu.k_0, \mu << 1$$
$$\Delta x \sim \Delta y \sim \Delta z \sim r_0; r_0 \Delta k \sim 1; t_0 \Delta\omega \sim 1; \qquad (5)$$
$$n_0.\omega_0/c = k_0; \quad \vec{E} = E_0.\vec{E}.$$

where $k_0$ and $\omega_0$ are the carrying wave number and carrying frequencies of the wave packet, $\Delta\omega$ and $\Delta k$ are the frequencies and wave number spectral bandwidth, and $t_0$, $r_0$ are the their temporal and the spatial dimension. We performed the order of magnitude analysis of (4) for localized wave packet bearing in mind relations (1) and (5). For dimensionless magnitudes we obtained:

$$\mu^2 k_0^2 \Delta\vec{E} \sim \mu^2 k_0^2 \frac{\partial^2 \vec{E}}{\partial t^2} \sim \mu^2 k_0^2 \frac{\partial^2 \vec{P}_{nl}}{\partial t^2}$$

These relations show as that all three terms are one order and for a strong field which satisfied (1) we must solve the nonlinear wave equation (4).

3. Separate of the variables and vortex soliton solution.

We have tried to solve (4) by using the method of separate of variables. The electric-field intensity vector $\vec{E}(\vec{r},t)$ is assumed to be:

$$\vec{E}(\vec{r},t) = (1/2i)\vec{E}(\vec{r})\exp(-i\omega t) - c.c.$$

so for stationary field we obtained a nonlinear Helmholtz type equation:

$$\Delta\vec{E} + k^2\vec{E} + n_2|\vec{E}|^2\vec{E} = 0 \qquad (6)$$

where $|\vec{k}|^2 = k^2 = \frac{n_0^2 \omega^2}{c^2}$ and $n_2 = k^2 \tilde{n}_2$.

In a Cartesian coordinate system the (6) is equal to a system of three nonlinear wave equations:

$$\frac{\partial^2 E_x}{\partial x^2} + \frac{\partial^2 E_x}{\partial y^2} + \frac{\partial^2 E_x}{\partial z^2} + k^2 E_x + n_2\left(|E_x|^2 + |E_y|^2 + |E_z|^2\right)E_x = 0$$
$$\frac{\partial^2 E_y}{\partial x^2} + \frac{\partial^2 E_y}{\partial y^2} + \frac{\partial^2 E_y}{\partial z^2} + k^2 E_y + n_2\left(|E_x|^2 + |E_y|^2 + |E_z|^2\right)E_y = 0 \qquad (7)$$
$$\frac{\partial^2 E_z}{\partial x^2} + \frac{\partial^2 E_z}{\partial y^2} + \frac{\partial^2 E_z}{\partial z^2} + k^2 E_z + n_2\left(|E_x|^2 + |E_y|^2 + |E_z|^2\right)E_z = 0$$

As (7) is a scalar system, it must be written in spherical coordinates.

$$\frac{1}{r}\frac{\partial^2 (rE_x)}{\partial r^2} + \frac{1}{r^2 \sin\theta}\frac{\partial}{\partial\theta}\left(\sin\theta \frac{\partial E_x}{\partial\theta}\right) + \frac{1}{r^2 \sin^2\theta}\frac{\partial^2 E_x}{\partial\varphi^2} + k^2 E_x + n_2\left(|E_x|^2 + |E_y|^2 + |E_z|^2\right)E_x = 0$$
$$\frac{1}{r}\frac{\partial^2 (rE_y)}{\partial r^2} + \frac{1}{r^2 \sin\theta}\frac{\partial}{\partial\theta}\left(\sin\theta \frac{\partial E_y}{\partial\theta}\right) + \frac{1}{r^2 \sin^2\theta}\frac{\partial^2 E_y}{\partial\varphi^2} + k^2 E_y + n_2\left(|E_x|^2 + |E_y|^2 + |E_z|^2\right)E_y = 0 \qquad (8)$$
$$\frac{1}{r}\frac{\partial^2 (rE_z)}{\partial r^2} + \frac{1}{r^2 \sin\theta}\frac{\partial}{\partial\theta}\left(\sin\theta \frac{\partial E_z}{\partial\theta}\right) + \frac{1}{r^2 \sin^2\theta}\frac{\partial^2 E_z}{\partial\varphi^2} + k^2 E_z + n_2\left(|E_x|^2 + |E_y|^2 + |E_z|^2\right)E_z = 0$$

Let us note here, that we kept the Cartesian basis for the basic vectors $\vec{x}, \vec{y}, \vec{z}$. We represented the components of the field as a product of radial and an angle part:

$$E_i = R(r)Y_i(\theta,\varphi), \quad i = x, y, z, \qquad (9)$$

with the additional condition for the angle part:

$$|Y_x(\theta,\varphi)|^2 + |Y_y(\theta,\varphi)|^2 + |Y_z(\theta,\varphi)|^2 = c = \text{const.} \quad (10)$$

Multiplying of each equation of (8) with corresponding $\dfrac{r^2}{R \cdot Y_i}$, $i = x, y, z$, and bearing in mind condition (10) we obtained:

$$\frac{r^2 \nabla_r^2 R}{R} + r^2\left(k^2 + c \cdot n_2 |R|^2\right) = -\frac{\nabla_{\theta,\varphi}^2 Y_i}{Y_i} = l(l+1) \qquad i = x, y, z \quad (11)$$

where l is a number and

$$\nabla_r^2 = \frac{1}{r^2}\frac{\partial}{\partial r}\left(r^2 \frac{\partial}{\partial r}\right) \quad (12)$$

$$\nabla_{\theta,\varphi}^2 = \frac{1}{\sin\theta}\frac{\partial}{\partial\theta}\left(\sin\theta \frac{\partial}{\partial\theta}\right) + \frac{1}{\sin^2\theta}\frac{\partial^2}{\partial\varphi^2}. \quad (13)$$

are corresponding the radial and the angular operators. In this way for the radial and the angular parts of waves functions there was obtained the equations:

$$\nabla_r^2 R + k^2 R + c n_2 |R|^2 R - \frac{l(l+1)}{r^2} R = 0 \quad (14)$$

$$\nabla_{\theta,\varphi}^2 Y_i + l(l+1) Y_i = 0 \quad (15)$$

The substitution (9) with condition (10) follow as to separate the nonlinear part in the radial component of the fields and for the angles components we have a usual linear eigenvalue problem. The equations (15) are well known equations and each of them has exact solutions of kind:

$$Y_i = Y_l^m(\theta,\varphi) = \Theta_l^m \Phi_m = \sqrt{\frac{2l+1}{4\pi}\frac{(l-m)!}{(l+m)!}} P_l^m(\cos\theta) e^{im\varphi} \qquad i = x, y, z. \quad (16)$$

where $P_l^m$ are the Legendre's functions for discrete series of numbers:

$l = 0,1,2.....; m = 0, \pm 1, \pm 2,........$ and $|m| \le 1$.

Going back to set (6), we can see that it is only possible to separate the variables for the spherical function, which satisfies condition (10). It is no hard to see that condition (10) is satisfied only by eigenfunctions with $l = 1, m = 0, \pm 1$, namely:

$$Y_1^{-1} = -\sqrt{\frac{3}{8\pi}} e^{-i\varphi} \sin\theta;$$
$$Y_1^0 = \sqrt{\frac{3}{4\pi}} \cos\theta; \quad (17)$$
$$Y_1^1 = \sqrt{\frac{3}{8\pi}} e^{i\varphi} \sin\theta;$$

For them there are three configurations:

$$|Y_1^{-1}|^2 + |Y_1^0|^2 + |Y_1^1|^2 = \frac{3}{8\pi}\sin^2\theta + \frac{3}{4\pi}\cos^2\theta + \frac{3}{8\pi}\sin^2\theta = \frac{3}{2\pi} = c = \text{const.},$$

$$|Y_1^1|^2 + |Y_1^0|^2 + |Y_1^1|^2 = \frac{3}{8\pi}\sin^2\theta + \frac{3}{4\pi}\cos^2\theta + \frac{3}{8\pi}\sin^2\theta = \frac{3}{2\pi} = c = \text{const.} \quad \text{and} \quad (18)$$

$$|Y_1^{-1}|^2 + |Y_1^0|^2 + |Y_1^{-1}|^2 = \frac{3}{8\pi}\sin^2\theta + \frac{3}{4\pi}\cos^2\theta + \frac{3}{8\pi}\sin^2\theta = \frac{3}{2\pi} = c = \text{const.}$$

By choosing for each field components one of these configurations, (10) is satisfied, and we see that the eigenfunctions (17) are solutions for the angular part of equation (6). The radial part of equation (14) has localized "de Broglie solitons" [11,13] solutions:

$$R = \frac{1}{2i}\frac{\sqrt{2}}{\sqrt{c \cdot n_2}}\left(\frac{e^{ikr}}{r} - \text{c.c.}\right)$$

When l=0, equation (14) is practically the radial part of the scalar stationary nonlinear Schredinger equation. It has been investigated in many earlier papers, but the existence of exact soliton solutions has only been found for a special kind of nonlinearity [12,13]. In this paper, it was found the existence of the localized vortex soliton solution for vector wave equation with eigenrotation momentum l=1, and for a fixed choose of the coordinates it is:

$$\vec{E} = -\frac{\sqrt{2}}{2i}\sqrt{\frac{3}{8\pi c n_2}}\left(\frac{e^{i(kr-\omega t)}}{r}e^{-i\varphi} - c.c.\right)\sin\theta\vec{x} + \frac{\sqrt{2}}{2i}\sqrt{\frac{3}{8\pi c n_2}}\left(\frac{e^{i(kr-\omega t)}}{r}e^{i\varphi} - c.c.\right)\sin\theta\vec{y} + \frac{\sqrt{2}}{2i}\sqrt{\frac{3}{4\pi c n_2}}\left(\frac{e^{i(kr-\omega t)}}{r} - c.c.\right)\cos\theta\vec{z}$$

(19)

The module of this vector is:

$$|\vec{E}| = \sqrt{|\vec{E}|^2} = \sqrt{\frac{2}{\tilde{n}_2}}\frac{\sin kr}{kr}$$

(20)

From (20) is it seen that the soliton solution is localized and has finite value when $r \to 0$. A known result from the scalar theory [2] is that in the case when the initial amplitude of the localized wave $R_0 > 1$, a self-focussing mode exists for the scalar field. For the amplitude of solution of the vector filed (18), we also obtained that $R_0 > 1$. But there is one additional possibility to compensate self-focussing of the field namely, by one rotation with orbital momentum l=1, that is a possibility to obtain stable vortex optical solitons. A full analysis of the stability of the vector soliton solutions of set (5) using the Hamiltonian formalism and other invariant, will be made in a next paper.

4. 3D+1 vector Schredinger equation.

Formally, we can made the substitution $k^2 E_i(r) = E_i(r)e^{ik^2 t} = -i\frac{\partial E_i(r,t)}{\partial t}$ and then the (6) become:

$$-i\frac{\partial E_x}{\partial t} + \frac{\partial^2 E_x}{\partial x^2} + \frac{\partial^2 E_x}{\partial y^2} + \frac{\partial^2 E_x}{\partial z^2} + n_2\left(|E_x|^2 + |E_y|^2 + |E_z|^2\right)E_x = 0$$
$$-i\frac{\partial E_y}{\partial t} + \frac{\partial^2 E_y}{\partial x^2} + \frac{\partial^2 E_y}{\partial y^2} + \frac{\partial^2 E_y}{\partial z^2} + n_2\left(|E_x|^2 + |E_y|^2 + |E_z|^2\right)E_y = 0$$
$$-i\frac{\partial E_z}{\partial t} + \frac{\partial^2 E_z}{\partial x^2} + \frac{\partial^2 E_z}{\partial y^2} + \frac{\partial^2 E_z}{\partial z^2} + n_2\left(|E_x|^2 + |E_y|^2 + |E_z|^2\right)E_z = 0$$

(21)

We thus obtain a time dependent vector nonlinear Schredinger equation. Using the same procedure as in the previous paragraphs, we obtained the same vortex soliton solutions (19) with the same eigenrotation momentum l=1. This is a generalization of the known result of the linear theory, for the mathematical identity between linear scalar Helmholtz and linear scalar Schredinger equations with respect to the stationary solutions.

5. Conclusion.

An order of magnitude analysis was made of the nonlinear wave equations in strong field approximation. Using the method of separation of variables, it is shown that exact localized vortex solitary soliton solutions exist with eigenrotation momentum l=1. The same soliton solutions exist also for the vector 3D+1 Schredinger equation. In contrast to the standard linear scalar theories where, the stationary solutions must be obtained for a infinite discrete series of numbers $k^2, l = 0,1,2,...m = 0,\pm 1,\pm 2,.....$, the nonlinear vector wave and Schredinger equations admit finite number of soliton solutions, because of the requirement the angular parts of equations to satisfy addition relations.

6. Acknowleglments: I am grateful to Stoil Donev for his comments on the manuscript.